# Science Literacy: Generative AI as Enabler of Coherence in the Teaching, Learning, and Assessment of Scientific Knowledge and Reasoning

Xiaoming Zhai[1], James W. Pellegrino[2], Matias Rojas[1], Jongchan Park[1], Matthew Nyaaba[1], Clayton Cohn[3], Gautam Biswas[3]

[1]*University of Georgia*

[2]*University of Illinois Chicago*

[3]*Vanderbilt University*

This chapter examines the potential of generative AI in enhancing science literacy across the K-16+ grade span, including its benefits as well as the conceptual and practical challenges that doing so presents. It begins with a discussion of what defines science literacy in the era of AI, including how AI has changed science and the demand for future citizens to be scientifically literate when AI is applied in their careers and lives. The chapter further discusses why science literacy presents such a challenge in K-16+ educational settings. It then develops an argument for the type of architecture needed for AI to assist in solving the problem by bringing coherence to the teaching, learning, and assessment of science knowledge and reasoning. Components of this architecture are illustrated with respect to the AI tools and capabilities needed for design and implementation. The chapter concludes with a consideration of what has been learned regarding both science literacy and AI, as well as what remains to be learned, including the research and development (R&D) needed, and the generalizability of this science literacy case to other disciplinary learning and knowledge domains.

## 1. Introduction

Many countries across the globe have long recognized that investments in public education can contribute to the common good, enhance national prosperity, and support stable families, neighborhoods, and communities. As argued in multiple position papers and policy reports, current economic, environmental, and social challenges point to the fact that education is even more critical today than it has been in the past (Shi, 2025b). Today's children can meet future challenges if they have opportunities to prepare for their future roles as citizens, employees, managers, parents, volunteers, and entrepreneurs. To achieve their full potential as adults, young people will need to acquire the full range of skills and knowledge that facilitate mastery of multiple academic subjects including science, mathematics, language arts, social studies, civics, and history. They will need to learn in ways that support not only retention, but also the use and application of skills and knowledge in the workplace and adult life.

In parallel developments regarding educational goals and policies, the last two decades have seen substantial and growing interest in changing the landscape of education through ideas labeled as "deeper learning" and "21st century skills." This global trend is indicative of a long-standing concern in education about the difficult task of equipping individuals with transferable knowledge and skills (Bellanca, 2014; Pellegrino & Hilton, 2012). Much of the discussion of deeper learning and 21st century skills have been couched in terms of broad, transversal competencies using generic labels such as problem solving, critical thinking, creativity, collaboration etc. (Foster & Piacentini, 2025). Arguments have been made, however, that deeper learning and the development of 21st century competencies do not happen separately from mastering disciplinary content and practices that are important to the discipline. Rather, deeper learning is the product of interconnected cognitive, interpersonal, and intrapersonal processes that enable students to thoroughly understand disciplinary content and recognize when, how, and why to apply that knowledge to solve new problems and engage in processes of critical thinking (Pellegrino & Hilton, 2012; Troitschankaia et al., 2025).

Perhaps it is not surprising then that concepts related to deeper learning and 21st century skills have come to be reflected in disciplinary frameworks and standards introduced over the last 15 years for Mathematics, Science, and Literacy in the United States (Pellegrino & Hilton, 2012), as well as other countries across the globe (OECD, 2023). For example, in the United States both contemporary mathematics and science standards include an emphasis on students using and applying knowledge in the context of disciplinary practices – that is, the actual everyday ways of knowing and doing that mathematicians and scientists employ in their respective fields. The basic premise for incorporating disciplinary practices into instruction is that learners, much like professionals, are more likely to advance or deepen their understanding when they have opportunities to use and apply knowledge to solve problems, reason with evidence, and/or make sense of phenomena. Contemporary Mathematics and Science frameworks and standards have in turn served as a source of inspiration for recently developed disciplinary standards and "big ideas" for newer, emergent K-16+ STEM subjects such as Computer Science, including Artificial Intelligence.

Contemporary STEM standards have profound implications for what constitutes literacy in each STEM discipline, how that develops over time with instruction, and what constitutes evidence of proficiency. An ongoing challenge is how to design curricular materials to support acquisition of these important competencies and how to organize classroom instruction, including the design and use of assessment, to promote student attainment of the complex disciplinary objectives and critical thinking skills embodied by contemporary STEM standards (Harris et al., 2023). The implications for what to teach and assess and how to do so at the classroom level are profound.

Juxtaposed with the conceptual and pragmatic developments mentioned above is the explosion of knowledge about AI and its applications, including the incredible opportunities that exist, as well as concerns for its impact, both positive and negative, on all facets of education (Allen & Kendeou, 2024). It has been hailed as both an amplifier for outcomes of the teaching and learning process, as well as a negative influence on what and how students learn and their ability to engage in critical thinking. Thus, in addition to thoughtful discussions and illustrations of its potential application in areas like STEM education (Zhai & Krajcik, 2024), there are numerous papers bemoaning its impact on the teaching, learning, and assessment process, across the K-16+ educational range (Guo et al., 2024). Embedded in this discourse are well-founded concerns about ethical and appropriate uses of AI tools and methods (Akgün & Krajcik, 2024).

The goal of this chapter is to examine and illustrate the potential of generative AI in enhancing science literacy across the K-16+ grade span, including its benefits, as well as the conceptual and practical challenges presented by attempting to design and implement AI supported systems that promote coherence among curriculum, instruction, and assessment in the science classroom. While we focus on the K-12 grade range to illustrate our arguments, we believe the implications extend well beyond those educational levels into multiple areas and levels of postsecondary STEM education.

In the next section, we explore current issues in science education and start by defining Science Literacy based on current science education frameworks and standards. The following section, Section 3, describes a human-in-the-loop (HITL) architecture to guide AI development and implementation in science, an architecture which we believe extends to other disciplinary areas as well. Section 4 outlines strategies for achieving coherence in teaching, learning, and assessment in science and provides illustrative examples. Finally, Section 5 concludes the chapter by discussing research, development, and implementation challenges that are key to advancing the use of generative AI with the goal of improving coherence in teaching, learning, and assessment in K-12+ science education, including its relevance to other educational fields.

## 2. Science Literacy in the Era of AI

### *2.1. What Do Students Need to Know and Be Able to Do?*

Science literacy has been described in various ways over time but in all cases it has focused on what students need to know and be able to do and the contexts in which we would expect proficiency to be demonstrated. Liu and Tripp (2025) describe three related but somewhat different visions of science literacy that have evolved over time. Vision I science literacy is based on what is required to be competent in science fields and careers, while Vision II science literacy is based on what people need to know and can do with science in relation to everyday life and society, regardless of whether they are working in science fields or not. Vision I science literacy is about mastery of science knowledge and processes, and Vision II science literacy is the everyday, contextual applications of science in society. Liu and Tripp (2025) note that Visions I and II can be thought of as two ends of a science literacy continuum (Roberts & Bybee, 2014). They argue that there is a Vision III that goes beyond Visions I and II and subsumes them given its focus on knowledge and action at the level of community and society.

All visions of Science Literacy have been influenced by serious concerns about what individuals need to know and be able to do that have emerged over the last two decades. Those concerns have arisen in part from student performance on tasks that mimic aspects of scientific investigations (Snyder & Dillow, 2013). Two unique types of activity-based tasks were administered as part of the 2009 U.S. National Assessment of Educational Progress (NAEP) science assessment (NCES, 2012; Zhai & Pellegrino, 2023). In addition to more typical paper-and-pencil questions, 4th, 8th, and 12th graders also completed hands-on and interactive computer tasks designed to reveal not just what students know at a factual level, but also how well they are able to reason through complex problems and apply science to real-life situations. While performing the interactive computer and hands-on tasks, students were asked to manipulate objects and perform actual experiments; these assessment tasks generated rich data on how students respond to scientific challenges. The 2012 NCES report of findings from these assessment tasks states that students succeeded in parts of investigations that involved limited data sets and making simple observations from that data. However, they faced difficulties with parts of investigations that had more variables to manipulate or required strategic decision-making to gather appropriate data. The NCES also noted that many students could draw correct conclusions from an investigation but were unable to explain their results.

  Such NAEP results, together with other findings in the science education literature, illustrate the disjuncture between students' knowledge of science facts and procedures, as assessed by certain types of science achievement tests, and their understanding of how that knowledge can be applied through the practices of scientific reasoning and inquiry. Recognition of this science education problem can be found in reports spanning elementary, secondary, and postsecondary education (K-16+). Collectively these reports present a consistent description of the science proficiency desired in contrast to that which students typically demonstrate. They include NRC reports on K-8 science education in formal and informal learning environments (NRC, 2007; NRC 2009), the redesigned curriculum and assessment frameworks for Advanced Placement science courses (e.g., College Board, 2011a, b), and even revisions in the nature of the science knowledge required for entry to medical school and how it should be assessed on the Medical College Admissions Test (e.g., AAMC, 2012). Seldom has such a consistent message been sent across K-16+ education as to the need for substantial change in what we expect students to know and be able to do in science, how science should be taught, and how it should be assessed.

## *2.2. Contemporary Science Frameworks*

The foregoing concerns contributed to a major reconceptualization of the nature of science literacy that was most clearly expressed in the NRC report *A Framework for K-12 Science Education* (National Research Council, 2012b). The *Framework* articulates three interconnected dimensions of competence. The first of these dimensions is *Disciplinary Core Ideas*. In response to criticisms of U.S. science curricula being "a mile wide and an inch deep" (Schmidt et al., 2002), especially when compared to other countries, the *Framework* identified and emphasized a small set of core ideas in four areas: (a) life sciences, (b) physical sciences, (c) earth and space sciences, and (d) engineering, technology, and the application of science. In doing so, the Framework aimed to reduce the long and often disconnected list of factual knowledge that students typically had to memorize. For example, key concepts in physical sciences include energy and matter, while in life sciences, they encompass ecosystems and biological evolution. Students encounter these core ideas throughout their school years, gradually deepening their understanding and knowledge. The second dimension is *Crosscutting Concepts*. The *Framework* identifies seven such concepts that have importance across many science disciplines; examples include patterns, cause and effect, systems thinking, and stability and change. The third dimension is *Science and Engineering Practices*. Eight key practices are identified, including asking questions (for science) and defining problems (for engineering); planning and carrying out investigations; developing and using models; analyzing and interpreting data, and engaging in argument from evidence.

While the *Framework's* three dimensions are conceptually distinct, the vision is one of coordination in science and engineering education such that the three are integrated in the teaching, learning, and doing of science and engineering. By engaging in the practices of science and engineering, students gain new knowledge about the disciplinary core ideas and come to understand the nature of how scientific knowledge develops. It is not just the description of key elements of each of the three dimensions that matters in defining science literacy and proficiency; a central argument of the *Framework* is that the meaning of proficiency is realized through performance expectations describing what students at various levels of educational experience should know and be able to do.  These performance expectations integrate the three dimensions and move beyond the vague terms, such as "know" and "understand," often used in previous science standards documents to more specific expectations like "analyze", "compare", "predict", "explain", and "model," in which the practices of science are wrapped around and integrated with core content.  Finally, the *Framework* makes the case that proficiency and expertise develop over time and increase in sophistication and power as the product of coherent and integrated systems of Curriculum, Instruction, and Assessment.

The *Framework* uses the three dimensions—the practices, crosscutting concepts, and core ideas of science and engineering—to organize the content and sequence of learning. This three-part structure signaled an important evolutionary shift for science education and presented the primary challenge for the design of both instruction and assessment—finding a way to describe and capture students' developing proficiency along these intertwined dimensions. The *Framework* emphasizes that research indicates that learning about science and engineering

"involves integration of the knowledge of scientific explanations (i.e., content knowledge) and the practices needed to engage in scientific inquiry and engineering design" (pg. 11). Both practices and crosscutting concepts are envisaged as tools (skills and strategies) for addressing new problems that are equally important for students' science learning as the domain knowledge topics with which they are integrated. Students who experience the use of these tools in multiple contexts as they learn science are more likely to become flexible and effective users of them in new problem contexts.

The *Framework* also uses the logic of learning progressions to describe students' developing proficiency in these three intertwined domains in a coherent way across grades K through 12, noting that "If mastery of a core idea in a science discipline is the ultimate educational destination, then well-designed learning progressions provide a map of the routes that can be taken to reach that destination" (p. 26). The stress on learning progressions is supported by research on learning described in the 2007 NRC report *Taking Science to School* and in other documents (Alonzo & Gotwals, 2012; Corcoran et al., 2009; Jin et al., 2024). The *Framework* builds on the idea of a developmental progression of student understanding across the grades by specifying grade band end point targets at grades 2, 5, 8 and 12 for each component of each core idea. For the practices and crosscutting concepts, the *Framework* also provides sketches of possible progressions for learning each practice or concept but does not indicate the expectations at any grade level. The *Next Generation Science Standards* (NGSS Lead States, 2013), described below, built on these suggestions and developed tables that define what each practice might encompass at each grade level; the *NGSS* also define the expected uses of each crosscutting concept for students at each grade level.

*2.3. Standards and Assessments*

To support the approach to science learning described above, the *Framework* states that assessment tasks must be designed to gather evidence of students' ability to apply the practices and their understanding of the crosscutting concepts in the contexts of problems that also require them to draw on their understanding of specific disciplinary ideas. In developing the NGSS, Achieve[1] and its partners turned these guidelines into standards that are explained by descriptions of how students at each grade are expected to use both the practices and crosscutting concepts, as well as the knowledge they should have of the core ideas (NGSS Lead States, 2013).

    The NGSS standards appear as clusters of performance expectations related to a particular aspect of a core disciplinary idea. For the Middle School Physical Science DCI of *Structure and Properties of Matter*, there are three performance expectations one of which is MS-PS1-2: *Analyze and Interpret Data on the Properties of Substances Before and After the Substances Interact to Determine if a Chemical Reaction has Occurred*. As illustrated in this example, each performance expectation asks students to use a specific practice and a crosscutting concept in the context of a specific element of the disciplinary knowledge relevant to a particular aspect of the core idea. Across the set of such expectations at a given grade level, each practice and crosscutting concept appears in multiple standards.

    In contrast to science standards like the *NGSS* that call for the integration of science practices and content knowledge, the prior generation of U.S. science standards (Council et al., 1995; NRC 1996, 2000), as well as those of other countries, treated content and inquiry as largely separate strands of science learning, and assessments followed suit. In some respects, the

---

[1] Achieve (https://www.achieve.org/next-generation-science-standards) is an independent, nonpartisan, nonprofit education reform organization dedicated to working with states to raise academic standards and graduation outcomes.

form the standards took contributed to this separation: content standards stated what students should know, and inquiry standards stated what they should be able to do. Consequently, assessments separately measured the knowledge and inquiry practice components. Thus, the idea of an integrated, multidimensional science performance presents a very different way of thinking about science proficiency. Disciplinary core ideas and crosscutting concepts serve as thinking tools that work together with scientific and engineering practices to enable learners to solve problems, reason with evidence, and make sense of phenomena. Such a view of competence also signifies that measuring proficiency solely as the acquisition of core content knowledge or as the ability to engage in inquiry processes free of content knowledge is neither appropriate nor sufficient.

To illustrate the instructional, learning, and assessment challenges posed by the conception of literacy and proficiency embedded in the *Framework*, we can consider its projected endpoint for K-12 science education. By the end of 12$^{th}$ grade all students—not just those interested in pursuing science, engineering, or technology beyond high school—should have gained sufficient knowledge and understanding to (1) engage in public discussions of science-related issues (such as the challenges of generating sufficient energy, preventing and treating diseases, maintaining supplies of clean water and food, and addressing problems caused by global environmental change); (2) be critical consumers of scientific information related to their everyday lives; and (3) continue to learn about science throughout their lives. Students should come to appreciate that science as a discipline, and current scientific understanding of the world, are the result of hundreds of years of creative human endeavors (National Research Council, 2012a).

*2.4. Visions and Levels of Science Literacy*

The above discussion can be juxtaposed with the visions of science literacy mentioned earlier. Liu and Tripp (2025) argued that the learning expectations for K-12 science education embodied in the *Framework* and *Next Generation Science Standards* should be viewed as a balance between Vision I and Vision II, as well as a significant advancement relative to prior Vision 1 science education standards (Council et al., 1995; NRC 1996, 2000). They also noted that the literature related to both visions of science literacy has focused almost exclusively on individuals, i.e., science literacy as a personal attribute. In a 2016 report, the National Research Council committee proposed that science literacy could be demonstrated at three aggregation levels: individual, community, and society (Dibner & Snow, 2016). Community and societal science literacy, while including individual science literacy as their foundations, assume that science literacy is distributed among members and agents of community and society. The necessity of science literacy for democracy, civic engagement, culture, and economic development is more dependent on science literacy at the community and societal levels than at the individual level. One can argue that recent trends in public discourse and debate regarding scientific issues such as climate change, the environment, disease prevention via vaccination, and fluoridation of drinking water clearly illustrate the need for science literacy at the community and societal level. The 2016 NRC Committee argued that because addressing complex social, cultural, political and environmental issues requires people's agency and active engagement, this collective aspect of science literacy can be considered as part of Vision III science literacy.

Liu and Tripp (2025) propose that science literacy can be represented as levelled visions with increasing sophistication. Vision III encompasses or subsumes Vision II, which also subsumes Vision I. Across the visions, there is progression from lower levels to higher levels, from fundamental knowing (e.g., facts, concepts, procedures), to using (e.g., explaining, evaluating, reasoning with, applying), to acting (e.g., engaging in citizen science and socio-scientific issues) which includes designing, building, and innovating practical solutions, products, and systems that are beneficial to communities and society (McElhaney et al, 2023; Mormina, 2019). In this chapter, further consideration of science literacy will focus primarily on those elements of its description as found in the *Framework* and *NGSS*.

Before leaving this discussion of the elements of science literacy we would be remiss in not noting that the science and engineering practices as articulated in the *Framework* need to be expanded to include uses of AI since scientists and engineers now regularly engage in the use of AI for multiple aspects of their work (Herdiska & Zhai, 2024). More and more, AI tools are being developed and used by scientists and engineers to support complex data analysis and assist with their problem-solving efforts by enabling analyses that help understand complex phenomena and generate solutions to difficult problems with demonstrated breakthroughs (Jumper et al., 2021; Khaleel et al., 2023). Thus, the development of science literacy needs to encompass aspects of AI literacy, especially as AI tools and practices become integral to the science teaching and learning environment. We will return to a discussion of the implications of AI literacy as part of science literacy in the final section of this chapter after considering the infusion of AI into the teaching, learning, and assessment of science.

*2.5. Implementing the Vision: Challenges and Prospects for AI*

In the 12+ years since the Framework and NGSS were first introduced, numerous papers and investigations have described the challenges associated with implementing the vision of science literacy at the core of both documents, with many offering guidance on how to do so. Early on, many of the challenges associated with developing multidimensional science assessments for classroom and/or large-scale use were clearly discussed in the NRC Report *Developing Assessments for the Next Generation Science Standards* (Pellegrino et al., 2014). That report included an extensive discussion of the design challenges of developing assessments aligned with multidimensional performance expectations and the approaches needed to develop such assessments systematically. It also considered the forms of evidence that would be needed to establish their validity in specific contexts of use.

A subsequent NRC Report *Guide to Implementing the Next Generation Science Standards* clearly discussed implications of the *Framework's* vision for the nature of the classroom teaching and learning environment (National Research Council, 2015). The table 1, developed by NSTA, captures several dramatic shifts in the nature of the classroom teaching and learning environment outlined in the *Framework* and *NGSS*. Note especially the implications of the student activities for teachers with respect to organization, management, assessment, and overall support of their students' learning progress.

Table 1. A new vision for science education (source: NRC, 2025)

| Science Education Will Involve Less… | Science Education Will Involve More… |
|---|---|
| Rote memorization of facts and terminology | Facts and terminology as needed while developing explanations and designing solutions supported by evidence-based arguments and reasoning |
| Learning of ideas disconnected from questions about phenomena | Systems thinking and modeling to explain phenomena and to give a context for the ideas to be learned |
| Teachers providing information to the whole class | Students conducting investigations, solving problems, and engaging in discussions with teachers' guidance |
| Teachers posing questions with only one right answer | Students discussing open-ended questions that focus on the strength of the evidence used to generate claims |
| Students reading textbooks and answering questions at the end of the chapter | Students reading multiple sources, including science-related magazines and journal articles and web-based resources; students developing summaries of information |
| Pre-planned outcome for "cookbook" laboratories or hands-on activities | Multiple investigations driven by students' questions with a range of possible outcomes that collectively lead to a deeper understanding of established core scientific ideas |
| Worksheets | Student writing of journals, reports, posters, and media presentations that explain and argue |
| Oversimplification of activities for students who are perceived to be less able to do science and engineering | Provision of supports so that all students can engage productively in science and engineering practices |
| Science Education Will Involve Less… | Science Education Will Involve More… |
| Rote memorization of facts and terminology | Facts and terminology as needed while developing explanations and designing solutions supported by evidence-based arguments and reasoning |
| Learning of ideas disconnected from questions about phenomena | Systems thinking and modeling to explain phenomena and to give a context for the ideas to be learned |
| Teachers providing information to the whole class | Students conducting investigations, solving problems, and engaging in discussions with teachers' guidance |

| Teachers posing questions with only one right answer | Students discussing open-ended questions that focus on the strength of the evidence used to generate claims |
| Students reading textbooks and answering questions at the end of the chapter | Students reading multiple sources, including science-related magazines and journal articles and web-based resources; students developing summaries of information |
| Pre-planned outcome for "cookbook" laboratories or hands-on activities | Multiple investigations driven by students' questions with a range of possible outcomes that collectively lead to a deeper understanding of established core scientific ideas |
| Worksheets | Student writing of journals, reports, posters, and media presentations that explain and argue |
| Oversimplification of activities for students who are perceived to be less able to do science and engineering | Provision of supports so that all students can engage productively in science and engineering practices |

It should come as no surprise, then, that progress in science teaching, learning, and assessment has proven challenging and slow since the introduction of the *Framework* and *NGSS*. Among the many challenges encountered was the slow emergence of curriculum materials with a strong alignment to the *Framework* and *NGSS* conceptions of multidimensionality. Included among these concerns was the lack of assessments requiring evidence of multidimensional proficiency and suitable for classroom or large-scale state use. Professional learning for teachers regarding understanding of each of the three dimensions associated with science literacy, especially the science and engineering practices, and their integration in instruction and assessment, has continued to pose a substantial challenge at all grade levels across K-12. The paucity of adequate professional learning opportunities for K-5 teachers has been especially noted (Pellegrino, 2021; Zhai & Crippen, 2025). Fortunately, substantial progress has been made in the current decade on multiple fronts with respect to instructional and assessment materials aligned with the literacy vision embodied in the *Framework* and *NGSS* (Harris et al., 2023).

One of the more significant developments in recent years is the effort to leverage AI capabilities to support various aspects of classroom instructional design and delivery, including effective assessment strategies. Zhai and Crippen (2025) have provided a concise summary of the types of work that have been done and are currently underway. Their analysis of the current situation in science education underscores the capabilities of AI for further progress and transformation of science education in the service of the development of science literacy.

> *"Positioned as a catalyst for innovation, AI introduces a suite of tools and frameworks that directly confront long-standing challenges in science education. These technologies include AI-supported scoring of open-ended written explanations and arguments or drawn models that delivers rubric-aligned scores and formative feedback at scale (Zhai., 2024; Cohn, et al, 2024), allowing students to practice modeling, explaining, and arguing from evidence while teachers track growth with reliability approaching human ratings. Adaptive learning platforms that tailor instruction to individual students' cognitive and emotional profiles, promoting differentiated learning experiences that foster deeper understanding (Gao & Zhang, 2025). Intelligent Tutoring Systems (ITS), such as AutoTutor and EER-Tutor, replicate aspects of one-on-one instruction by offering real-time, customized feedback based on student inputs, thereby enhancing engagement and* *learning efficiency (Latif et al., 2025). Similarly, Open-Ended Learning Environments (OELEs), such as Betty's Brain (Biswas et al, 2016) and Meta Tutor (Azevedo et al, 2022), use AI tools that systematically monitor students' learning and problem-solving strategies and provide metacognitive support, thereby promoting critical thinking skills that improve science learning. Generative AI further enables automatic creation of instructional materials, such as lesson plans and assessments (Lee & Zhai, 2024), reducing teachers' routine workload and freeing them to concentrate on higher-order instructional tasks that more directly support science learning. In addition, AI-supported learning analytics can provide educators with fine-grained, real-time insights into student behavior, comprehension levels, and progression, enabling data-informed instructional decisions for science learning (Gašević et al., 2022; Sajja et al., 2025). These analytics help identify learning bottlenecks, predict at-risk students, and support timely interventions."* (Zhai & Crippen, 2025)

While many of the affordances and capabilities of AI have begun to infiltrate K-12 science education, it must be noted that they are primarily focused on specific issues associated with implementing the instructional and assessment practices envisioned by the *Framework.* They provide excellent proof-of-concept examples regarding how to design and enact aspects of learning environments consistent with the vision of teaching, learning, and assessment articulated in the *Framework,* including the types of classroom activities outlined in the Table 1 shown earlier. We believe that to fully capitalize on the power and affordances of AI in supporting coherence among curriculum, instruction, and assessment in the science education classroom, a comprehensive framework for AI development and implementation is needed to guide future progress. In the next Section, we outline key elements of such an architecture, including arguments for their relevance. In Section 4, we then describe progress in designing and implementing critical aspects of the architectural design.

**3. A Human-in-the-Loop Framework to Guide AI Development and Implementation**

As generative AI continues to reshape the scientific literature, it becomes imperative to conceptualize an architecture and infrastructure that ensures AI technologies augment, rather than replace, human agency in teaching, learning, and assessment. This section presents a human-in-the-loop (HITL) framework designed to support the coherent integration of AI systems into science education. The framework positions educators, students, and AI systems within an interconnected ecosystem where human judgment, ethical reflection, and contextual understanding guide the development and deployment of AI tools (Mosqueira-Rey et al., 2023).

*3.1. Conceptual Foundations of a Human-in-the-Loop Framework*

The HITL framework asserts that the advancement of AI in science education must be grounded in an epistemologically balanced relationship between human expertise and algorithmic computation. It contends that effective science education cannot rely on automation alone but must sustain iterative cycles of human design, supervision, interpretation, and refinement of AI-mediated processes. This position challenges techno-deterministic assumptions that portray AI as a self-sufficient agent, arguing instead that its educational value emerges only through human mediation. By framing AI as a cognitive and pedagogical collaborator rather than a replacement for human intellect, the framework advances the claim that educators' interpretive judgment, disciplinary knowledge, and ethical reasoning are indispensable for ensuring that AI tools genuinely support the cultivation of scientific literacy. Such a stance situates the educator not as a passive user of AI outputs but as an active co-constructor of learning environments that harness

AI to deepen reasoning, personalize instruction, and make the assessment of complex scientific outcomes more authentic and meaningful.

At its core, the HITL framework builds upon three interdependent principles that directly support the development of science literacy—the ability to apply scientific knowledge, reasoning, and inquiry to real-world contexts (see Figure 1):

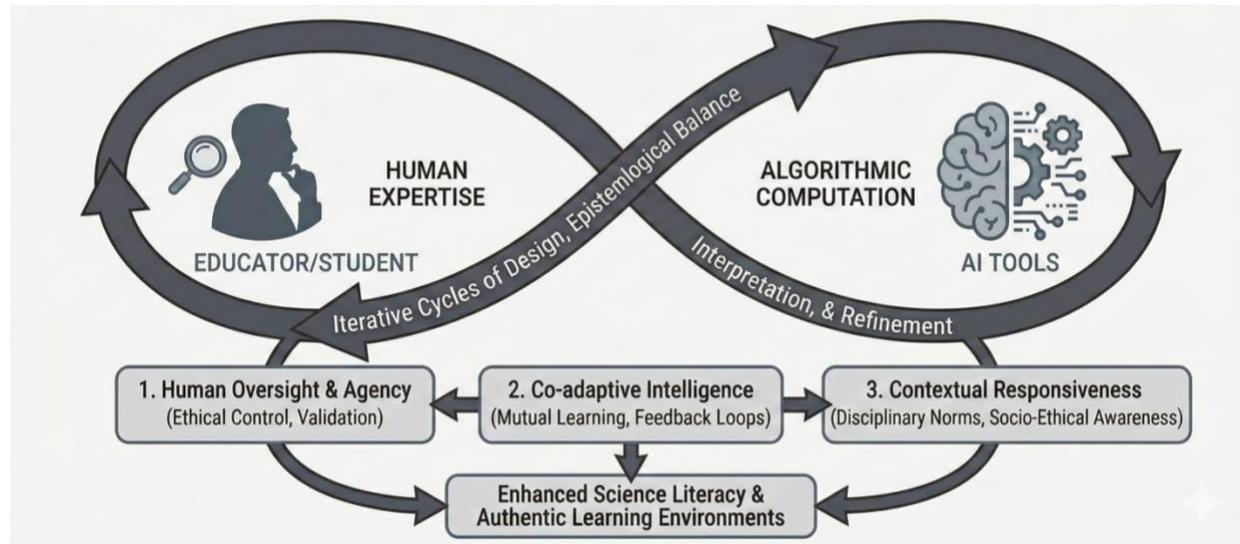

Figure 1. Human-in-the-Loop framework for AI integration in science education

***Human Oversight and Agency.*** AI systems should remain under human interpretive and ethical control, ensuring that automated processes align with educational values. In the context of science literacy, this principle reinforces the importance of human judgment in validating evidence, interpreting uncertainty, and distinguishing correlation from causation—skills central to scientific reasoning. Human oversight safeguards the epistemic standards of science, reminding learners that knowledge construction is a human endeavor grounded in critical inquiry.

***Co-adaptive Intelligence.*** Human users and AI tools should evolve together through feedback loops, promoting mutual learning and refinement. When teachers and students iteratively interact with AI systems, they co-construct understanding and refine their scientific reasoning. For instance, a teacher may use AI-generated hypotheses or models as catalysts for discussion, while students critique, modify, and extend them. This reciprocal adaptation models the iterative nature of scientific inquiry itself, where evidence, interpretation, and theory evolve in dialogue.

*Contextual Responsiveness.* AI applications must be sensitive to disciplinary norms, learner diversity, and the socio-ethical implications of scientific knowledge. Science literacy demands not only conceptual understanding but also awareness of how science operates within cultural, environmental, and ethical contexts. A contextually responsive AI system can tailor examples, data, and explanations to diverse learners and local phenomena, helping students see science as a living, human practice with social consequences.

These principles collectively ensure that AI enhances the authenticity, inclusiveness, and integrity of science literacy by aligning technological affordances with the human capacities for reasoning, reflection, and ethical awareness.

*3.2. Architectural Layers of the AI-Human Ecosystem*

A robust architecture for generative AI integration in science education can be conceptualized through four interlocking layers: (1) data infrastructure, (2) model development, (3) pedagogical interface, and (4) human governance. Each of these layers is theoretically grounded in established frameworks from the learning sciences, sociotechnical systems theory, and AI ethics. The *data infrastructure* layer aligns with constructivist and sociocultural perspectives (Bransford et al., 2000; Vygotsky & Cole, 1978), which emphasize that scientific knowledge emerges from shared tools and representations; thus, data must reflect authentic social and epistemic practices. The *model development* layer draws on theories of co-regulation and human–machine symbiosis (Hollnagel & Woods, 2005), positioning AI as an augmentative partner that adapts to human cognitive processes. Principles of cognitive apprenticeship justify the *pedagogical interface* layer (Collins et al., 1991) and distributed cognition (Rogers, 1997), recognizing that meaningful learning occurs when humans interact with intelligent systems in context-rich environments. Finally, the *human governance* layer draws on insights from the ethical AI and participatory design literatures (Floridi et al., 2021), underscoring the necessity of human accountability, transparency, and democratic oversight in educational technology ecosystems. Collectively, these theoretical foundations justify the multilayered architecture as essential for sustaining human-centered science literacy in the age of generative AI.

Figure 2 illustrates a GenAI approach that adopts an HITL approach to developing a pedagogical agent that supports students during their inquiry and problem-solving processes when learning a science topic. A human-in-the-loop GenAI pedagogical agent, developed by Cohn and the OELE Lab at Vanderbilt, uses a theory-driven Evidence–Decision–Feedback (EDF) architecture to support students' inquiry and problem-solving in science (Cohn et al., 2026, in press). EDF integrates foundational constructs from well-established theories in education and the learning sciences. The Evidence module gathers and interprets signals from students' actions and dialogue to update a learner model. Following Evidence-Centered Design (ECD; Mislevy & Hartel, 2006), domain models specify competencies; evidence models identify observable indicators, and task models define activities that elicit them. Stealth assessment (Shute, 2011) enables continuous inference, producing diagnostics of mastery, strategy use, problem-solving progress, and motivation for the Decision module. The Decision module translates diagnostics into instructional intent, guided by the Zone of Proximal Development (Vygotsky, 1978) and Social Cognitive Theory (Bandura, 2001) to infer needs and select appropriate behavioral, cognitive, and metacognitive strategies to motivate students and support their learning. The Feedback module converts intent into concrete utterances and actions. Grounded in social constructivism (Adams, 2006), it treats feedback as a mediational tool, drawing on dialogue policy, discourse history, context, and the learner model to deliver appropriate responses, manage sequencing, and guide learners toward more sophisticated problem-solving.

In this cyclic framework, student actions generate evidence that informs the agent's decisions, which are then delivered as targeted feedback; subsequent student responses enter the loop as new evidence, enabling continual, theory-grounded adaptivity and personalization. The design mandates traceability to facilitate a researcher-teacher-student HITL approach to the design and development of the GenAI agent. In other words, the pedagogical agent feedback must be explicitly linked to the Decision module's dialogue policy, and that policy must be grounded in inferences from the Evidence module. Classroom teachers and educators work closely with AI researchers to develop the three EDF modules, and the human-readable chain of reasoning allows teachers to interpret and trust the system's actions, helps students understand the rationale for scaffolds and prompts, and enables researchers to diagnose, evaluate, and refine the agent's adaptive behavior. A case study of this iterative HITL approach with a high school teacher designing formative feedback for students working on a science modeling task demonstrated an iterative loop: the GenAI module helped the teacher identify students' difficulty thresholds and intervention points, and subsequent teacher insights—including the need for a deliberate lag before intervention—informed improvements to the feedback system (Cohn et al., 2024).

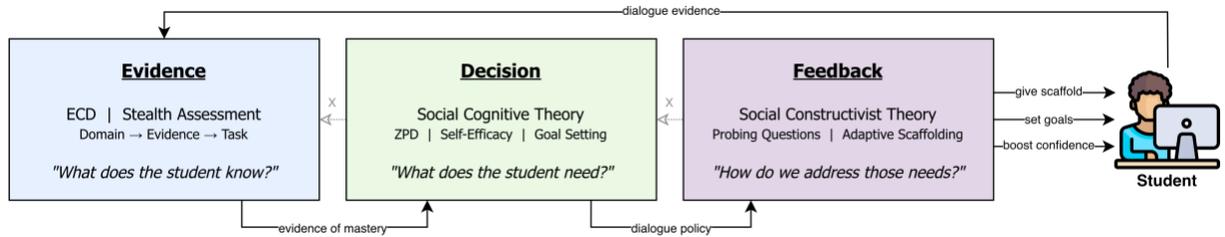

Figure 2. An Example of a HITL-designed GenAI Agent for adaptive scaffolding and feedback.

Building on the EDF framework, recent work advances a user-centered, human-in-the-loop partnership among researchers, teachers, and students to design a semi-autonomous collaborator agent for the C2STEM (Collaborative Computational STEM) learning environment, where learners co-construct computational models of scientific processes. The approach, illustrated in Figure 3, implements a multi-agent GenAI system comprising four interlinked sub-agents—*StrategyAgent, KnowledgeAgent, AssessmentAgent*, and *DialogAgent*—coordinated through a unified learner model. This model is continuously updated using environment logs, students' collaborative dialogue, and chat histories with the LLM-powered collaborator. The multi-agent framework further enhances the HITL approach by enabling teachers to work closely with AI researchers to guide the development of the different components of the GenAI agent. The result is a coherent, adaptive system that aligns with teacher-supported pedagogy with real-time evidence. In a classroom deployment, the agent operated effectively and received strong endorsements from both students and teachers (Cohn et al., 2025).

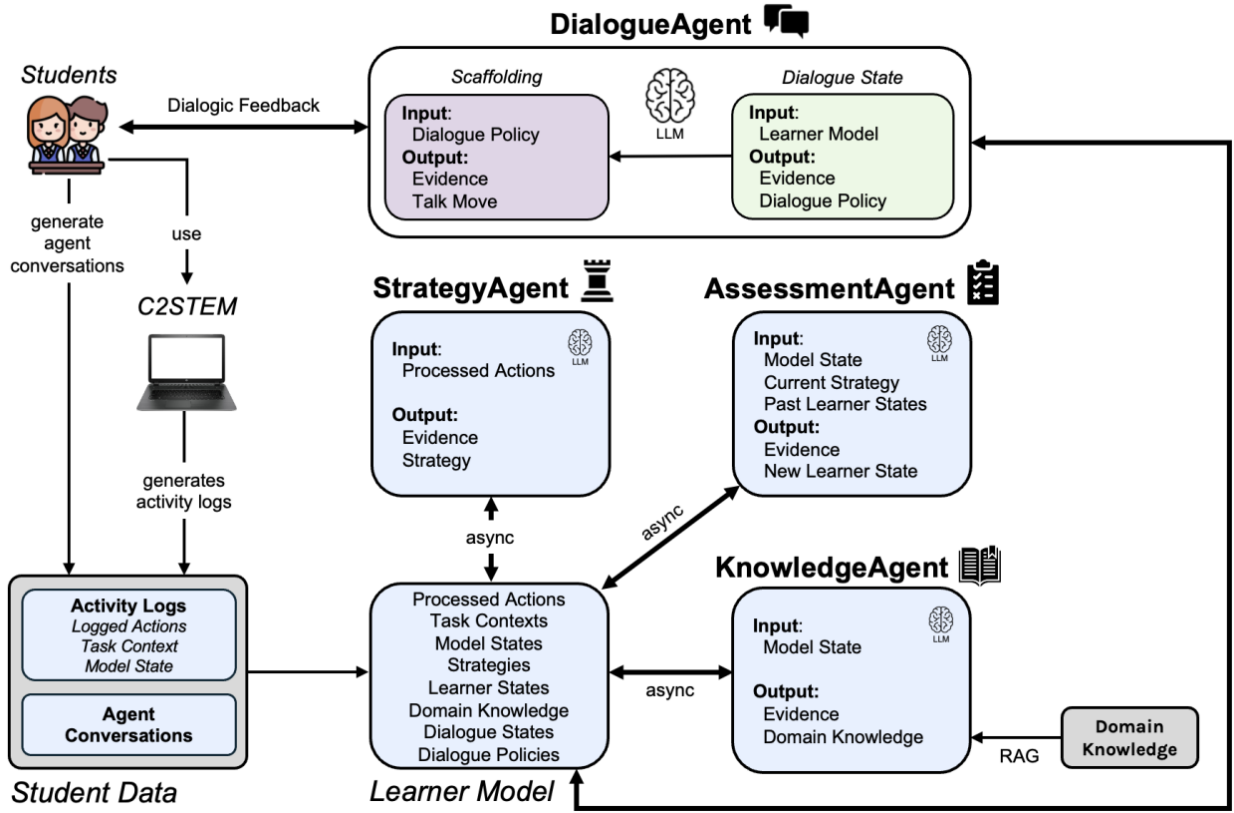

Figure 3: The LLM-based GenAI agent with four sub-agents: *StrategyAgent, KnowledgeAgent, AssessmentAgent*, and *DialogAgent*. Gray boxes correspond to data stores. Blue, green, and purple boxes correspond to Evidence, Decision, and Feedback modules—respectively.

*3.2.1. Data Infrastructure Layer*

This foundational layer governs the sourcing, structuring, and stewardship of data used to train, fine-tune, and operate AI models. It is crucial to establish epistemic credibility within educational AI ecosystems. Informed by sociocultural learning theory (Vygotsky, 1978) and critical data studies (Boyd & Crawford, 2012), it recognizes data not as a neutral substrate but as a socially constructed entity shaped by cultural assumptions, epistemic values, and power relations. Consequently, the development of responsible data infrastructure must prioritize not only technical efficiency but also democratic accountability, ethical inclusivity, and epistemic integrity to support science literacy.

*Ethical data curation*, guided by frameworks of data justice (Heeks & Renken, 2018) and inclusive epistemology (Harding, 1991), emphasizes the deliberate incorporation of diverse epistemic traditions and underrepresented voices into datasets used to train AI systems. This ensures that AI-mediated science education reflects pluralistic forms of knowledge and resists the perpetuation of systemic bias. Privacy and security measures, informed by digital ethics and theories of contextual integrity (Floridi et al., 2021) must move beyond procedural compliance to embrace relational ethics—treating learners' data as expressions of personhood and intellectual agency. Protecting these data through encryption, anonymization, and transparent consent practices upholds learners' epistemic autonomy and reinforces trust in educational AI systems.

Equally important is *semantic alignment*, which draws on semantic web theory (Berners-Lee et al., 2006) and authentic science learning frameworks (Chinn & Malhotra, 2002). This principle requires that curated datasets embody the authentic logic of scientific inquiry—such as hypothesis generation, empirical testing, and evidence-based argumentation—so that AI models accurately mirror the cognitive and methodological dimensions of science. By aligning data representation with disciplinary epistemology, the infrastructure layer enables AI to engage learners in higher-order reasoning rather than superficial recall. Ultimately, this layer serves as the epistemic backbone of AI systems, grounding their operations in valid, equitable, and context-sensitive knowledge representations that transform raw data into a meaningful foundation for human–AI collaboration in science education.

### 3.2.2. Model Development Layer

This layer concerns the design, training, fine-tuning, and iterative refinement of generative models that underpin science learning within the HITL framework (Mittal et al., 2024). Rather than viewing model development as a purely technical enterprise, this stage constitutes an epistemic process in which computational systems are designed to complement and extend human reasoning. Drawing from socio-technical systems theory (Trist, 1981), design-based research traditions (Brown et al., 1989), and cognitive systems engineering (Hollnagel & Woods, 2005), the model development layer highlights the importance of maintaining human interpretive control throughout the lifecycle of AI model creation and deployment.

Educators play a central role in model design, serving as epistemic co-constructors who shape the conceptual fidelity of AI outputs. Teacher-in-the-loop annotation processes ensure that datasets and outputs align with disciplinary reasoning and curricular goals, grounding models in authentic scientific inquiry (Collins et al., 1991; Lave & Wenger, 1991). Through iterative collaboration between educators, researchers, developers, and learners, generative models are continuously fine-tuned to reflect evolving educational objectives, consistent with formative assessment principles (Black & Wiliam, 2009) and theories of co-adaptive intelligence (Licklider, 2008). This co-design process not only enhances model reliability and transparency but also supports the development of explainable AI systems that enable teachers and learners to trace reasoning steps behind generated outputs (Doshi-Velez & Kim, 2017; Wu et al., 2025).

Moreover, this layer integrates insights from epistemic cognition research (Chinn et al., 2011) to ensure that AI systems reflect the logic, uncertainty, and iterative nature of scientific reasoning. Transparent and interpretable model architectures allow educators to engage critically with AI outputs, verifying their validity and relevance to scientific understanding. When effectively implemented, this integration bridges computational intelligence with human epistemology, positioning AI as a reflective collaborator in science learning rather than an autonomous decision-maker. In doing so, the model development layer ensures that AI remains accountable to human expertise, pedagogical integrity, and the epistemic standards of the scientific enterprise.

*3.2.3. Pedagogical Interface Layer*

This layer concerns the design, training, and continual refinement of generative models that underpin science learning. It represents not only a technical domain but an epistemically charged process through which AI systems are constructed to reflect, mediate, and extend human reasoning. Within a HITL framework, this phase embeds human expertise at every stage of development, ensuring that AI systems serve pedagogical and ethical ends rather than mere computational optimization.

Drawing on socio-technical systems theory (Trist, 1981) and design-based research traditions (The Design-Based Research Collective, 2003), this layer underscores the necessity of co-evolution between human and machine intelligence. Educational AI cannot be static or self-sufficient; its design must remain open to interpretive feedback and evolve in tandem with curricular goals, disciplinary epistemologies, and social contexts. As scholars such as Suchman (2007) and Hollnagel and Woods (2005) have argued that meaningful human–machine collaboration depends on constant calibration of responsibility and control, ensuring that systems remain both intelligible and pedagogically trustworthy.

Teacher participation constitutes the epistemic core of this layer. Building on participatory design scholarship (Simonsen & Robertson, 2013) and situated learning theory (Lave & Wenger, 1991), teachers act not as passive annotators but as co-designers and validators of AI behavior. Their engagement ensures that generative models internalize the conceptual and procedural dimensions of scientific reasoning—such as hypothesis formation, data interpretation, and argument construction—reflecting authentic practices of inquiry. Teachers' reflective judgment thus mediates between algorithmic suggestion and scientific credibility, safeguarding the integrity of educational outcomes.

Model adaptation, grounded in formative assessment principles (Black & Wiliam, 2009) and dynamic systems theory (Thelen & Smith, 1994), further reinforces the dialogic nature of human–AI co-evolution. Through iterative feedback loops among educators, learners, and developers, models are fine-tuned to reflect evolving pedagogical insights and learning trajectories. This iterative responsiveness mirrors Vygotsky's (1978) notion of mediated learning, where tools adapt to learners' zones of proximal development, aligning AI's cognitive scaffolding with human developmental logic.

Equally critical is the pursuit of transparency and interpretability. Research in explainable AI (Doshi-Velez & Kim, 2017; Miller, 2019) and epistemic cognition (Chinn, Buckland, & Samarapungavan, 2011) demonstrates that intelligibility is not an aesthetic preference but a precondition for epistemic trust. By providing visible reasoning traces and contextual explanations, interpretable AI systems enable educators and learners to interrogate claims, identify inconsistencies, and evaluate the coherence of scientific arguments. This process transforms AI from a black-box instrument into a pedagogical partner that fosters metacognitive awareness and epistemic reflection.

In conclusion, when guided by informed oversight and reflective reasoning, AI becomes not merely a technological artifact but an epistemic collaborator capable of advancing scientific literacy through authentic engagement with the uncertainty, creativity, and logic of scientific inquiry.

*3.2.4. Human Governance Layer*

At this layer, generative AI functions as an embedded learning partner within instructional workflows, shaping how students, teachers, and intelligent systems co-construct knowledge. The pedagogical interface represents the visible dimension of the human–AI partnership—where theoretical aspirations for collaboration, transparency, and epistemic agency manifest in practical teaching and learning contexts. Drawing upon sociocultural learning theory (Vygotsky, 1978), distributed cognition (Rogers, 1997), and cognitive apprenticeship models (Collins, Brown, & Newman, 1989), this layer operationalizes learning as a dialogic and mediated process in which AI becomes an active participant in the scaffolding of scientific understanding.

In interactive co-reasoning environments, generative AI supports learners in constructing explanations, visualizing phenomena, and simulating scientific processes. This aligns with constructivist perspectives that emphasize inquiry, hypothesis generation, and reflective discourse as key drivers of scientific literacy (Bruner, 1961). AI systems designed for co-reasoning can act as cognitive amplifiers, offering dynamic representations and hypothetical scenarios that challenge students to reason through complexity. However, the pedagogical effectiveness of such environments depends on continuous teacher mediation—teachers must monitor, interpret, and contextualize AI contributions to maintain epistemic rigor and avoid overreliance on algorithmic authority.

The interface also facilitates personalized feedback loops, a critical mechanism for adaptive learning. Drawing from formative assessment research (Black & Wiliam, 2009) and intelligent tutoring systems literature (Anderson et al., 1995; Biswas et al., 2016), AI can generate nuanced, individualized feedback that diagnoses misconceptions, recommends strategies, and supports metacognitive reflection (Basu et al., 2017; Munshi et al., 2023). However, such personalization must remain interpretively grounded: educators evaluate AI-generated insights within the broader trajectories of student development, ensuring that feedback advances conceptual growth rather than mere performance optimization. In this sense, personalization becomes a co-regulated process, guided by human judgment and pedagogical values (Cohn et al., 2024).

Moreover, generative AI transforms the teacher's role as an instructional designer through collaborative authoring tools. Building on theories of teacher agency and knowledge co-design (Koehler & Mishra, 2005; Penuel et al., 2007), educators can employ AI systems to co-create inquiry materials, simulations, or assessment tasks that integrate content knowledge with practices of scientific reasoning. This collaboration expands teachers' creative capacity, allowing them to focus on higher-order instructional goals such as fostering argumentation, ethical reasoning, and interdisciplinary synthesis. Importantly, such tools exemplify the notion of teachers as epistemic designers—professionals who shape how knowledge is represented, mediated, and learned in the digital age. Additionally, teachers can utilize LLM-based tools to create agents that score and explain short-answer formative assessments in K-12 science, guided by teachers' rubrics (Cohn et al., 2024). They can also provide explanations of the scores, allowing students to receive timely feedback and enhance their learning.

Ultimately, the human–AI interactions sustain the authenticity of scientific inquiry while promoting active learning, reflection, and agency. By grounding AI design in learning theory and empirical research, this layer reframes technology not as an instructional substitute but as a medium for cultivating the intellectual virtues central to science literacy—curiosity, critical evaluation, and the pursuit of coherent understanding.

# 4. Implementation Pathways: Building Coherence Across Teaching, Learning, and Assessment

To operationalize the HITL framework, implementation must unfold across three interdependent dimensions—teaching, learning, and assessment—each governed by principles of coherence, transparency, and reflective inquiry.

## *4.1. Teaching: AI as a Co-Designer of Learning Experiences*

Within the domain of teaching, the integration of generative AI demands a reconceptualization of instructional design as a process of co-authorship between educators and intelligent systems. Grounded in theories of technological pedagogical content knowledge (Koehler & Mishra, 2005), cognitive apprenticeship (Collins, Brown, & Newman, 1989), and teacher agency (Biesta et al., 2015; Zhai, 2024), this dimension views educators as epistemic designers who collaborate with AI to curate, interpret, and adapt AI-generated resources to promote scientific reasoning and inquiry.

For example, Nyaaba and Zhai (2025) developed the Culturally Responsive Lesson Planner (CRLP-GPT) powered by an Interactive Semi-Automated (ISA) prompt architecture (see Figure 4). Unlike standard GPT prompting, which produces generic instructional outputs, the ISAP workflow engages teachers in a structured dialog in which the AI asks targeted, context- and culture-specific questions, and teachers respond with their content knowledge, classroom expertise, and understanding of students' cultural assets. This iterative questioning process activates key cognitive skills, including pedagogical decision-making, cultural reasoning, scientific sensemaking, and anticipatory judgment, as teachers refine the AI's drafts toward contextually aligned instructional materials. AI then serves as a cognitive partner that expands teachers' capacity to design rich, inquiry-driven learning experiences. Rather than producing a finished product, the AI supports teachers by generating pedagogical suggestions, curricular maps, or simulations informed by learner data, thereby enhancing the personalization and responsiveness of science instruction. Teachers, in turn, evaluate AI's outputs, identify inaccuracies or cultural mismatches, request revisions, and integrate local knowledge systems into the design (Nyaaba & Zhai, 2025).

As argued by Selwyn (2019) and Luckin (2018), this augmentation does not diminish the centrality of teacher judgment. Instead, the ISA workflow reinforces teachers' roles as developers, evaluators, and meaning-makers, ensuring that AI outputs are grounded in classroom, community, and scientific realities (Nyaaba & Zhai, 2025). Such contextual adaptation aligns with situated cognition theory (Lave & Wenger, 1991) and HITL, indicating that AI gains significance only through its application in authentic, socially embedded practices overseen by teachers.

Furthermore, the ethical dimension of AI-assisted teaching necessitates deliberate reflexivity. Teachers must interrogate not only the utility but also the epistemic assumptions and potential biases embedded in AI systems. Drawing on critical digital pedagogy (Morris & Stommel, 2018) and AI ethics education (Jobin et al., 2019), educators should cultivate discussions that invite students to examine the social, cultural, and moral implications of AI in science. Such engagement transforms classrooms into spaces of ethical reasoning, fostering students' understanding of science as both a cognitive and moral enterprise. As illustrated by Nyaaba and Zhai (2025), the cognitive demands placed on teachers throughout this co-design process, critiquing AI drafts, generating cultural examples, evaluating scientific coherence, adjusting difficulty levels, and confirming conceptual accuracy, demonstrate that AI-assisted teaching is not a passive activity. It is an active, reflective, and iterative form of pedagogical reasoning that strengthens teacher agency rather than weakening it.

Ultimately, the teaching dimension of the HITL framework reaffirms that while generative AI can extend the scope of instructional creativity and precision, it must remain subordinate to human interpretation, judgment, and ethical stewardship. In this configuration, teachers are not displaced by technology but empowered as co-designers of learning experiences that advance scientific literacy through inquiry, reflection, and critical engagement with AI.

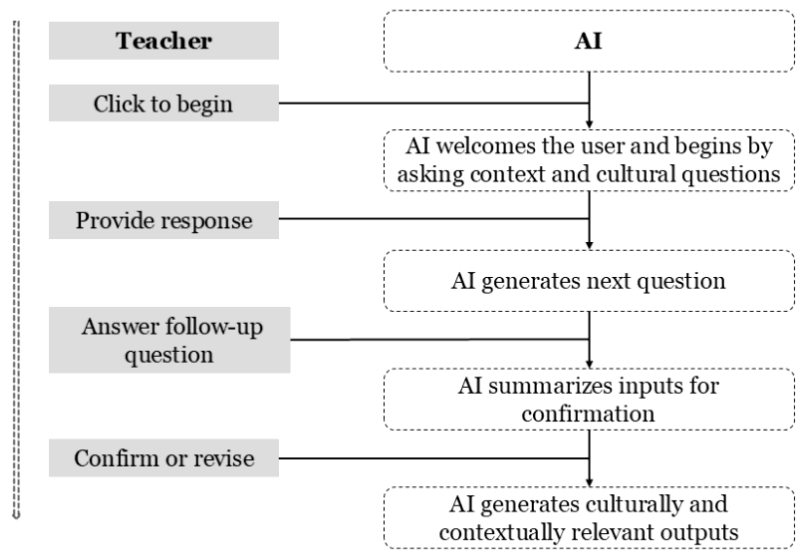

Figure 4. Interactive Semi-Automated Prompt Co-Design Workflow Between Teachers and AI (Adapted from Nyaaba and Zhai, 2025)

### *4.2. Learning: AI as a Scaffold for Inquiry and Reflection*

For learners, AI systems act as powerful mediators of reasoning, inquiry, and conceptual growth, especially when they are designed to keep students "in the loop" as active sense-makers rather than passive recipients of answers, promoting sustained engagement with scientific knowledge and authentic disciplinary practices. Drawing on constructivist and sociocultural theories (Vygotsky, 1978; Driver et al., 1994; Bransford, Brown, & Cocking, 2000), AI technologies scaffold understanding by helping students visualize abstract scientific relationships, coordinate multiple representations of data and ideas, model argumentative reasoning, and engage in structured metacognitive reflection. In this way, AI supports learners' transition from novice to expert through guided participation and the co-construction of meaning.

In the GENIUS project, we have developed an inquiryAgent to facilitate AI-based scientific inquiry, aiming to improve students' science discipline-based AI literacy (Zhai, 2025). Students use AI to extend their inquiries while retaining control over which questions they pursue, which evidence they consider, and how far their conclusions can be generalized (Herdliska & Zhai, 2024). Mirroring how contemporary scientists employ AI tools across the research cycle, classroom uses of AI typically fall into a small set of bounded roles that remain anchored in human judgment: as a computational instrument, AI functions like a powerful but transparent tool for computing, visualizing, or organizing information whose inputs, settings, and outputs students must specify and interpret; as a generative collaborator, it helps explore literature or prior data, surface patterns, and propose hypotheses, plans, or lines of critique that students must evaluate and revise; and as a proactive assistant, it can monitor data or student work and highlight surprising patterns or suggest next steps, but its recommendations are always open to human scrutiny, negotiation, and, when appropriate, rejection.

Generative and predictive AI tools facilitate multimodal exploration of scientific phenomena by integrating linguistic, visual, and symbolic representations that foster systems thinking (Jacobson & Wilensky, 2006) and conceptual coherence, especially when embedded in inquiry workflows that foreground evidence rather than producing only fluent text. When incorporated in dialogic learning environments, conversational AI systems function as inquiry partners that are tightly coupled to students' ongoing investigations (for example, data analysis, modeling, or simulation activities), sustaining socratic exchanges that stimulate hypothesis formation, evidence evaluation, and epistemic debate while helping learners refine questions, plan simple experiments or simulations, analyze patterns in data, and connect results back to underlying mechanisms. Such dialogic engagement echoes traditions of guided discovery and cognitive apprenticeship, situating learners within authentic scientific discourse communities where they practice reasoning and critique as scientists do. In the AI-based inquiry design, each phase of investigation yields small, scorable artifacts—such as brief evidence captions or decision logs—that make students' coordination of data, models or simulations, and claims visible for both learners and teachers and that can be instantiated in classroom tools combining AI with modeling workspaces or virtual labs, such as the example shown in Figure 5.

Figure 5. Students use GenAgent[2] train a model to differentiate two types of leaves, mimicking how scientists using AI to do research

Additional examples of such environments that support inquiry learning and computational scientific modeling include the C2STEM (Collaborative Computational STEM) learning environment (see Figure 6), where students build simulation models of physical phenomena (e.g., forces, energy) using domain-specific blocks that correspond to scientific quantities (e.g., force, mass, acceleration, velocity, displacement) (Hutchins et al., 2020). Students run simulations, observe behaviors, and iteratively improve their code to match model outcomes with target phenomena. Tasks are organized around ask–hypothesize–model–test–evaluate–revise. The system encourages learners to make predictions, design and conduct virtual experiments, gather and analyze data (tables/graphs), and justify revisions based on evidence.

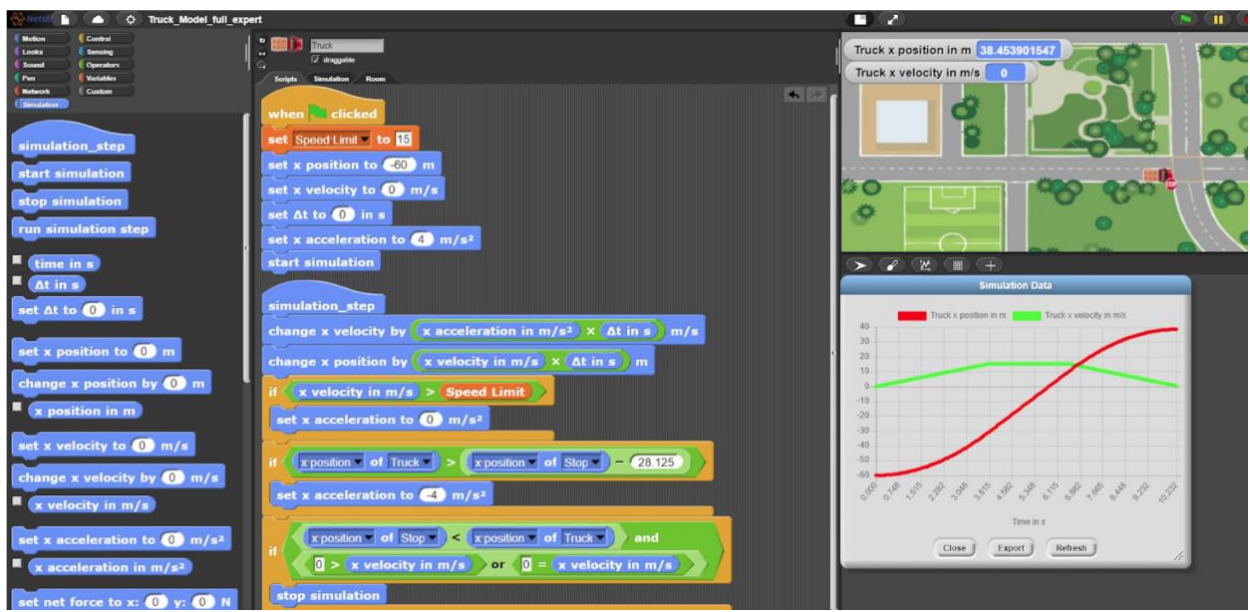

Figure 6: The C2STEM Model Building and Problem-Solving Interface

Furthermore, AI-generated feedback promotes metacognitive regulation and epistemic awareness by prompting students to evaluate the strength of their arguments, confront uncertainty, and revise misconceptions, especially when that feedback is explicitly grounded in the quality and limits of available evidence (e.g., by pointing to noisy measurements, model uncertainty, missing controls in a simulation, or underrepresented cases in a dataset). Aligned with theories of self-regulated learning (Zimmerman, 2002; Winne, 2013) and epistemic cognition (Chinn, Buckland, & Samarapungavan, 2011), these reflective processes help learners internalize scientific habits of mind—such as skepticism, curiosity, and intellectual humility. At the same time, AI-generated feedback can cultivate AI-specific literacies, such as tying claims to auditable artifacts (e.g., metrics, visualizations, or simulation logs), tracking the provenance of data and prompts that shape a model's or agent's behavior, and judging when AI support is sufficient or should be overridden—for example, when an AI suggestion conflicts with well-established domain knowledge or with safety and equity constraints. When integrated into thoughtfully designed instructional contexts, AI becomes not a substitute for human reasoning but a catalyst for inquiry-driven, ethically reflective science learning that deepens both conceptual understanding and scientific literacy, and offers students practice in collaborating with AI while retaining responsibility for scientific judgment and ethical use.

---

[2] GenAgent is an intelligent system developed in the National GENIUS Center, aiming to improve students' science discipline-based AI literacy.

A recent study introduced a human-in-the-loop AI framework analyzing self-regulated learning (SRL) and socially shared regulation of learning (SSRL) in an embodied, mixed-reality classroom (Fonteles et al., 2024; 2025 (in press)). Figure 7 shows middle school students enacting the photosynthesis process. A new codebook by learning science and AI researchers fine-tuned deep learning models processing multimodal data—movement, gestures, gaze, speech, and interactions. A late-fusion LLM combined signals to segment and classify SRL–SSRL behaviors, validated via an interactive timeline. During a two-day photosynthesis activity with four students, the system identified 470 segments in ~40 minutes, with Enacting (66.4%) and Interacting (15.5%) most common. Human evaluation identified challenges due to overlapping metacognitive behaviors (e.g., Interacting and Reflection, Enacting and Monitoring). This work demonstrated how AI-human integration enabled hybrid intelligence for classroom behavior analysis and suggested future improvements to behavioral definitions, data alignment, validation tools, and teacher-oriented summaries (Fonteles et al., 2026 (in press)).

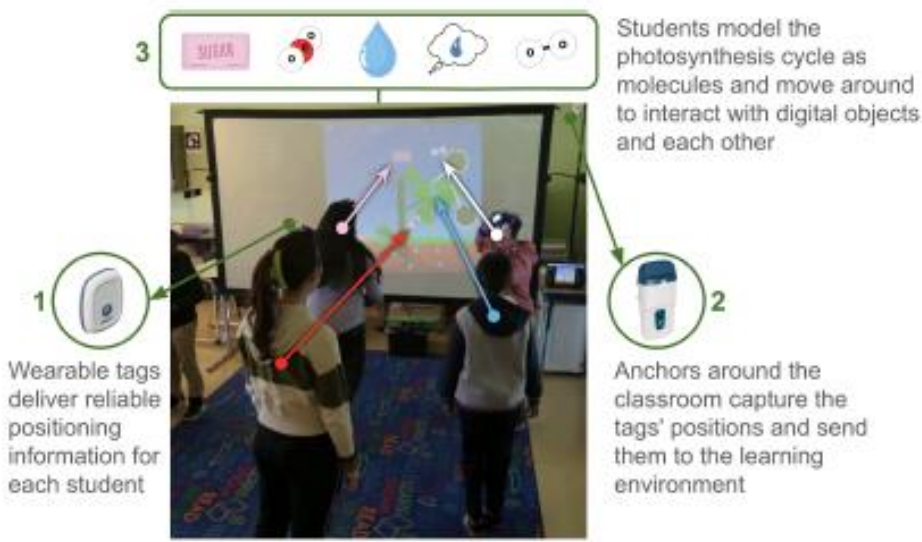

Figure 7. Real-time embodied visualization of the photosynthesis cycle using molecule avatars mapped to students' classroom movements

*4.3. Assessment: AI as a Partner in Evaluating Scientific Reasoning*

Generative AI possesses transformative potential to reimagine assessment practices, particularly within formative and performance-based contexts, when situated within the HITL framework. Informed by research on formative assessment (Black & Wiliam, 2009), learning analytics (Sajja et al., 2025; Selwyn, 2019), and authentic assessment theory (Gulikers et al., 2006), this approach positions AI not as a replacement for teacher judgment but as an analytical partner that enhances the interpretive and diagnostic dimensions of assessment. By analyzing patterns in students' explanations, laboratory reports, argumentation, and interaction traces from performance-based tasks, AI systems can identify conceptual trends, misconceptions, and reasoning strategies that may otherwise escape immediate human detection. However, teachers remain responsible for interpreting these outputs, ensuring that algorithmic evaluations align with contextual, ethical, and disciplinary standards.

Figure 8 illustrates how these capabilities are instantiated in the AI-Scorer system used for the Gas Filled Balloons performance-based task in a chemistry unit (Harris et al. 2023). In this example, the system provides real-time scores on five fine-grained, three-dimensional performance expectations that span disciplinary core ideas, science and engineering practices, and crosscutting concepts. The system aggregates these analyses into four teacher-facing dashboards: (a) average and individual student time on the item, (b) clusters of students with similar performance along a learning progression, (c) rubric-based scores for each expectation represented with color-coded indicators, and (d) a drop-down view of individual students' responses with associated scores and teacher comments or remarks. Together, these views help teachers rapidly identify emerging patterns in students' scientific reasoning and decide which groups or individuals require further discussion or support, while preserving teachers' epistemic responsibility for how the information is interpreted and acted upon.

Even when such dashboards are available, authenticity verification constitutes a crucial dimension of this model. Drawing on theories of situated learning (Lave & Wenger, 1991) and authentic scientific practice (Chinn & Malhotra, 2002), teachers must triangulate AI-generated assessments and automated summary reports of student performance with evidence from real-world inquiry tasks and their own first-hand observations of students. Through this synthesis, they judge the validity and epistemic fidelity of the AI-generated evidence. These judgments, in turn, inform revisions to tasks and pedagogical decisions for subsequent learning activities. In doing so, assessment becomes an iterative process of verification, reflection, and negotiation between computational inference and human expertise. This maintains the authenticity of scientific reasoning by ensuring that assessments capture not only outcomes but also processes of inquiry and understanding.

Moreover, learning analytics dashboards such as those in AI-Scorer, guided by research in educational data science (Ifenthaler & Yau, 2021) and metacognitive feedback (Winne et al., 2013), provide educators and learners with visualized insights into learning trajectories at individual, group, and classroom levels. For teachers, these synthesized dashboards support reflective practice by helping them track patterns over time and across tasks and make timely instructional adjustments. For students, the feedback visualizations and summaries make their progress and lingering difficulties more transparent, supporting metacognitive reflection and self-regulation. When governed by principles of interpretive control and data ethics (Williamson, 2017), such tools ensure that human actors maintain epistemic authority and responsibility. In this integrated configuration, assessment becomes a dialogic and dynamic process—one that merges computational precision with human discernment, fostering a culture of reflection, accountability, and authentic scientific reasoning.

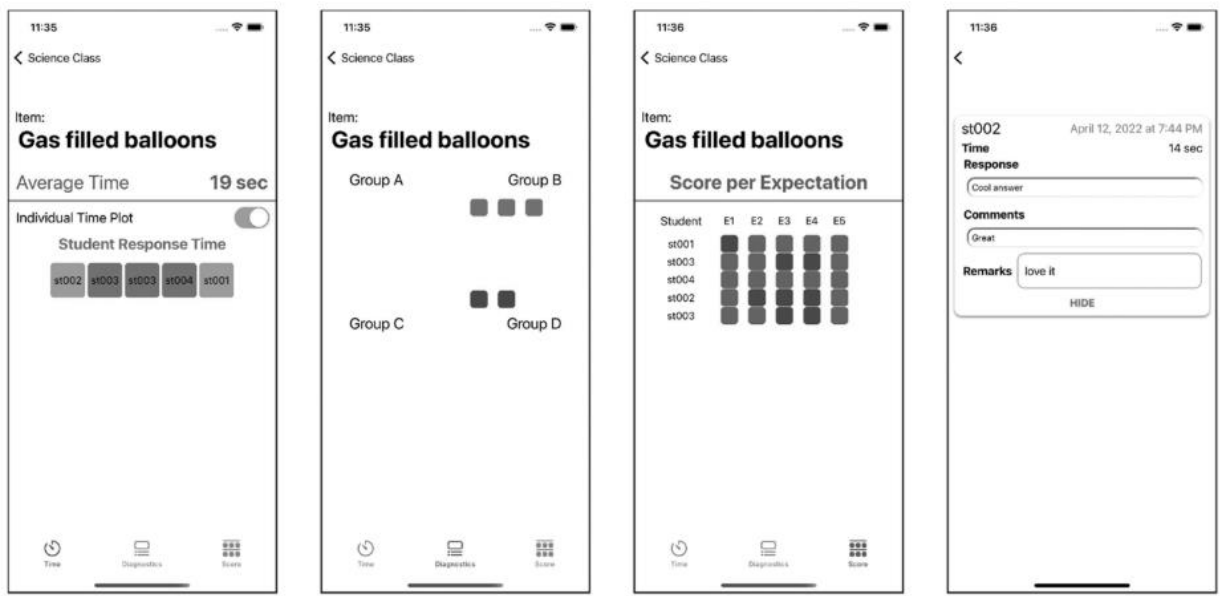

Figure 8. AI-Scorer system used to support teachers' instructional decision making in formative assessment practice (see Latif et al., 2024b)

## 5. Research, Development, and Implementation Issues

Advancing the coherent integration of generative AI into science education requires sustained research, long-term development efforts, and thoughtful implementation strategies that attend to the epistemic, pedagogical, ethical, and infrastructural complexities outlined throughout this chapter. Although the preceding sections have demonstrated the conceptual promise and emerging prototypes of human–AI collaboration in teaching, learning, and assessment, the realization of these possibilities at scale remains an ongoing challenge. The field stands at a critical juncture: existing innovations provide compelling proof-of-concept demonstrations, yet their widespread adoption demands robust evidence, technological reliability, and organizational capacity across K-16+ contexts. This final section outlines the major issues that must be addressed if generative AI is to fulfil its potential as a catalyst for science literacy and for coherence across curriculum, instruction, and assessment.

A central research need concerns understanding how AI-driven tools shape the development of scientific reasoning and inquiry practices over time. While early findings suggest that generative AI can scaffold complex reasoning, facilitate modeling, and support argumentation (Guo et al., 2024), the long-term effects of AI-augmented inquiry on students' epistemic agency remain insufficiently understood. In particular, more empirical work is needed to determine how learners negotiate the interplay between AI-generated suggestions and their own judgment, how they develop appropriate skepticism toward computational outputs, and how they cultivate the dispositions characteristic of scientific thinking, including curiosity, persistence, and epistemic humility, when working alongside intelligent systems (Crippen et al., 2025). Such research should address developmental trajectories across age groups and examine whether AI-supported inquiry differentially benefits students with varying linguistic, cultural, or cognitive backgrounds. Without this empirical grounding, efforts to embed AI into science curricula risk unintentionally reinforcing superficial forms of engagement or narrowing conceptions of scientific reasoning.

Another important and emerging body of work that directly intersects with these research challenges is the development of Discipline-Based Artificial Intelligence Literacy (DAIL) (Zhai, 2025). Coherent implementing AI in teaching, learning, and assessment demands DAIL for both students and teachers. DAIL provides a framework for conceptualizing how AI literacy must be contextualized within authentic disciplinary practices, moving beyond general-purpose AI knowledge to emphasize the integrated capabilities required to use and evaluate AI in the epistemic, methodological, and ethical contexts of specific fields (see Figure 9). DAIL includes six components: (1) Conceptual knowledge of AI in the discipline, (2) Practical skills in using AI tools, (3) Data and computational literacy within the discipline, (4) Critical evaluation and reasoning, (5) Ethical and societal understanding in the discipline, and (6) Collaborative and reflective disposition. DAIL closely aligns with the forms of reasoning and performance emphasized in multidimensional science learning, which underscores the need for assessments and instructional materials that embed AI within authentic disciplinary tasks, thereby reinforcing the coherence principles at the heart of the HITL framework. As systems aim to scale AI-supported learning environments, DAIL offers a valuable guide for specifying learning objectives, designing integrated learning progressions, and supporting teacher development to ensure that AI strengthens rather than fragments disciplinary learning.

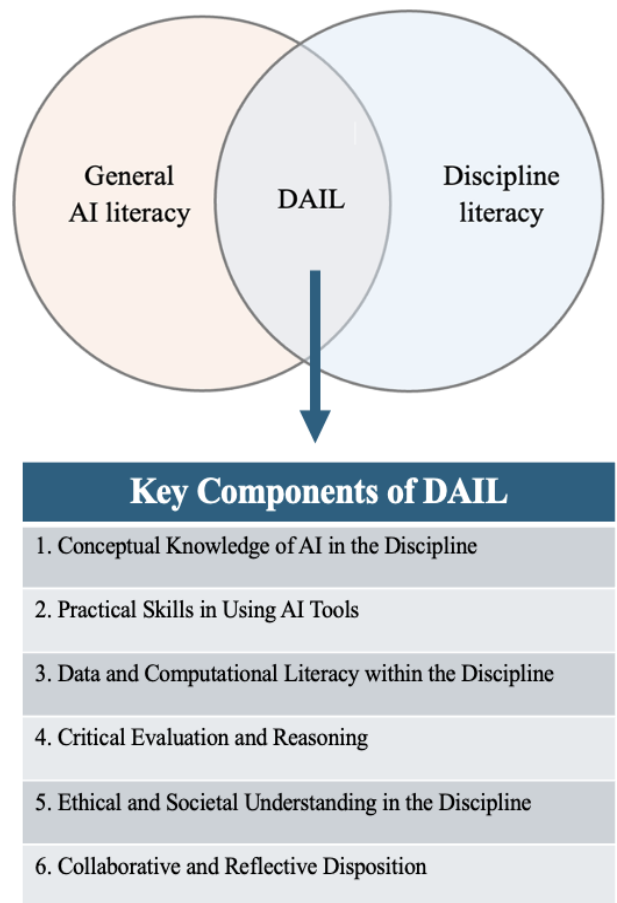

Figure 9. Disciplinary-based AI literacy and components (see Zhai, 2025)

Related to this is the challenge of systematically designing and validating AI systems that faithfully represent and support the multidimensional model of science proficiency articulated in the *Framework* (National Research Council, 2012b) and NGSS (NGSS Lead States, 2013). As emphasized earlier, the integration of disciplinary core ideas, crosscutting concepts, and science and engineering practices places substantial cognitive and representational demands on instructional and assessment systems. Generative AI has the potential to help address these demands, but only when systems are grounded in careful design principles, supported by high-quality data, and subjected to rigorous validation. At present, many AI tools rely on large, general-purpose models that lack fidelity to disciplinary reasoning or lack the grounding needed to produce evidence-based explanations and tasks (Latif et al., 2024a; Liu et al., 2023). Substantial developmental work is required to build models trained on scientifically authentic data representations, discipline-specific corpora, and annotated examples that reflect high-quality reasoning. Moreover, validation must go beyond predictive accuracy to evaluate epistemic fidelity: whether a system's explanations reflect legitimate scientific logic, whether feedback aligns with intended learning progressions, and whether generated tasks genuinely elicit multidimensional performances.

A further set of issues emerges around the sociotechnical infrastructure required to support implementation in real classrooms. Even when high-quality AI systems are available, their integration into everyday instruction hinges on teachers' capacity to use them productively and critically. This raises pressing questions about professional learning and teacher agency (Zhai, 2024). As demonstrated in the examples of ISA-based lesson planning and AI-assisted inquiry facilitation, teachers must engage in complex forms of pedagogical judgment, including evaluating the epistemic soundness of AI outputs, contextualizing tasks for their learners, mediating student, AI interactions, and attending to ethical considerations. Developing these competencies requires professional learning models that are sustained, practice-based, and designed to cultivate teachers' dual role as both users and co-designers of AI tools. Teacher preparation programs, certification requirements, and ongoing professional development must therefore evolve to incorporate AI literacy—not merely as technical training, but as a deep engagement with the epistemological, ethical, and pedagogical dimensions of human–AI collaboration (Nyaaba & Zhai, 2024; Shi, 2025a). Implementation research will be essential for understanding how teachers develop these competencies over time and how school systems can create supportive conditions for their enactment.

Ethical and governance concerns remain equally urgent. As highlighted in earlier sections, introducing AI into science education creates new responsibilities for data stewardship, algorithmic transparency, and equity (Akgün & Krajcik, 2024; Crippen et al., 2025). Research is needed to establish governance frameworks that ensure AI systems protect learner privacy, minimize bias, and operate in ways consistent with democratic values. These concerns extend to issues of access: without deliberate efforts to address resource disparities, AI-enhanced science learning could exacerbate existing educational inequalities, making advanced inquiry experiences available only to well-resourced schools. Implementation at scale must therefore be accompanied by policy commitments, funding structures, and technical supports that ensure equity of access and prevent the stratification of learning opportunities.

Finally, a broader question concerns the generalizability of science-focused AI architectures and practices to other disciplinary domains. While science provides a rich testbed given its emphasis on inquiry, modeling, and evidence-based reasoning, the principles outlined in the HITL framework have wider relevance. Disciplines such as mathematics, engineering, social studies, and language arts are increasingly encountering parallel challenges related to integrating complex disciplinary practices with conceptual understanding. Research is needed to examine how the architectural layers, co-adaptive learning processes, and participatory design approaches described here might be adapted to the epistemic structures and pedagogical goals of these other domains. Such comparative work would help determine whether the science literacy case represents a domain-specific pathway or a broader model for reimagining disciplinary learning in the age of AI.

In sum, the path forward requires coordinated progress across empirical research, technological development, policy design, and classroom implementation. The central promise of generative AI lies not in automating instruction or assessment but in enabling new forms of coherence—where curriculum, pedagogy, and evaluation mutually reinforce students' development of scientific literacy. Achieving this vision will require long-term inquiry into how students reason with AI, how teachers orchestrate human–AI partnerships, how systems can be designed to embody disciplinary epistemologies, and how educational institutions can sustain these innovations equitably. The work ahead is substantial, yet the potential rewards—a more coherent, inclusive, and inquiry-rich science education ecosystem—justify the ambition.

**Acknowledgments:** *The research reported here was supported by the Institute of Education Sciences, U.S. Department of Education, through Grant R305C240010 (PI Zhai). The opinions expressed are those of the authors and do not represent views of the Institute or the U.S. Department of Education. Additionally, Gautam Biswas was partially supported by the NSF AI Institute Engage AI grant #NSF-DRL-2112635.*

**Declaration:** *During the preparation of this work the author(s) used ChatGPT to check grammar and polish the wordings. After using this tool/service, the authors reviewed and edited the content as needed and take full responsibility for the content of the publication.*